\documentclass[acmsmall,screen,nonacm]{acmart}
\usepackage{algorithm}
\usepackage{algorithmic}

\acmJournal{PACMPL}
\acmNumber{OOPSLA}
\acmYear{2027}
\acmVolume{}
\acmArticle{}
\acmDOI{}
\acmISBN{}
\setcopyright{none}
\AddToHook{cmd/ps@firstpagestyle/after}{\fancyfoot{}}
\usepackage{amsmath}
\usepackage{graphicx}
\usepackage{booktabs,tabularx}
\usepackage{listings}
\usepackage{xcolor}

\microtypecontext{spacing=nonfrench}

\graphicspath{{figures/}}

\definecolor{CodeAccent}{RGB}{29,78,216}
\definecolor{PanelFrame}{RGB}{120,120,120}
\definecolor{PanelTitleBg}{RGB}{238,238,238}

\newcommand{\fpmpanelfont}{\fontsize{8}{9.5}\selectfont}

\lstdefinestyle{fpmpanel}{
    language=C,
    basicstyle=\fpmpanelfont\ttfamily\color{black},
    keywordstyle=\bfseries,
    commentstyle=\color{black},
    stringstyle=\color{black},
    emphstyle=\color{CodeAccent}\bfseries,
    numbers=none,
    frame=none,
    columns=fullflexible,
    keepspaces=true,
    showstringspaces=false,
    breaklines=false,
    tabsize=4,
    xleftmargin=0pt,
    xrightmargin=0pt,
    aboveskip=0pt,
    belowskip=0pt
}

\newsavebox{\fpmpanelboxA}
\newsavebox{\fpmpanelboxB}
\newlength{\fpmpanelheight}
\newlength{\fpmpanelpad}
\newcommand{\fpmhalfinner}{\dimexpr0.49\textwidth-1pt-2\fpmpanelpad\relax}
\newcommand{\fpmsplitinner}{\dimexpr(\textwidth-1pt-2\fpmpanelpad-13pt)/2\relax}

\newcommand{\fpmpanelfit}{%
  \setlength{\fpmpanelheight}{\dimexpr\ht\fpmpanelboxA+\dp\fpmpanelboxA\relax}%
  \ifdim\dimexpr\ht\fpmpanelboxB+\dp\fpmpanelboxB\relax>\fpmpanelheight
    \setlength{\fpmpanelheight}{\dimexpr\ht\fpmpanelboxB+\dp\fpmpanelboxB\relax}%
  \fi}

\newcommand{\fpmpanel}[3]{%
  \begingroup
  \setlength{\fboxsep}{0pt}\setlength{\fboxrule}{0.5pt}%
  \fcolorbox{PanelFrame}{white}{%
    \begin{minipage}[t]{\dimexpr#1-2\fboxrule\relax}%
      \colorbox{PanelTitleBg}{\makebox[\linewidth][l]{%
        \fpmpanelfont
        \rule[-0.4\baselineskip]{0pt}{1.3\baselineskip}%
        \hspace{\fpmpanelpad}\bfseries #2}}\par
      {\color{PanelFrame}\hrule height \fboxrule}%
      \vspace{3pt}%
      \hspace*{\fpmpanelpad}%
      \begin{minipage}[t][\fpmpanelheight][t]{\dimexpr\linewidth-2\fpmpanelpad\relax}%
        #3%
      \end{minipage}\par
      \vspace{3pt}%
    \end{minipage}}%
  \endgroup}

\newcommand{\fpmpanelpair}[2]{%
  \fpmpanelfit
  \noindent\fpmpanel{0.49\textwidth}{#1}{\usebox{\fpmpanelboxA}}\hfill
  \fpmpanel{0.49\textwidth}{#2}{\usebox{\fpmpanelboxB}}\par}

\newcommand{\fpmpanelsplit}[1]{%
  \fpmpanelfit
  \noindent\fpmpanel{\textwidth}{#1}{%
    \usebox{\fpmpanelboxA}\hfill
    {\color{PanelFrame}\vrule width 0.5pt}\hfill
    \usebox{\fpmpanelboxB}}\par}

\newcommand{\takeaway}[2]{\par\smallskip\noindent\fbox{\parbox{\dimexpr\linewidth-2\fboxsep-2\fboxrule\relax}{\textbf{Takeaway #1.} #2}}\par\smallskip}

\begin{document}

\title[VulnSAGE: Adjudicating SAST Reports]{\textsc{VulnSAGE}: Adjudicating SAST Reports by Connecting LLM Reasoning to Program Structure}

\author{Jinbao Chen}
\authornote{Both authors contributed equally to this research.}
\email{zkd18cjb@mail.ustc.edu.cn}
\author{Zixiang Zhou}
\authornotemark[1]
\email{zkd25zzx@mail.ustc.edu.cn}
\author{Yu Zhang}
\authornote{Corresponding author.}
\email{yuzhang@ustc.edu.cn}
\author{Yihong Wu}
\email{peterwuyihong@mail.ustc.edu.cn}
\affiliation{%
  \institution{University of Science and Technology of China}
  \country{China}}

\begin{abstract}
Static Application Security Testing (SAST) reports flag operations whose safety depends on project-specific context, and maintainers must decide which reports describe real defects. Large language models (LLMs) can interpret this context, but adjudicating a report with them requires the decisive code, which often lies far from the reported line, and a decision basis that program analysis can check. We present \textsc{VulnSAGE}, an adjudication system for SAST reports that connects LLM reasoning to program structure in both directions. Its report-centered slicer recovers the dependencies of the reported operation together with the program structure that explains them, such as the caller object on which a helper acts. CLAIM, an intermediate representation for report judgments, records each model interpretation as a semantic fact that cites the program relations it assumes. Program analysis checks these premises, and the verdict follows from the facts whose premises hold. Both mechanisms share a program graph, which lets the model cite the code it reads and lets the checker query those citations. Statically supported true positives then guide proof-of-concept generation and runtime reproduction. Across 187 labeled reports from four corpora covering five vulnerability families, \textsc{VulnSAGE} rejects 89 of 97 false alarms and retains 85 of 90 real defects. Under a common PoC procedure, 25 of its positive verdicts reproduce at runtime, compared with three for the LLM-based adjudicator LLM4SA. Applied to 45,440 alerts raised by five SAST tools on the 200 most-starred C/C++ projects on GitHub, it reproduces 104 defects at runtime. Maintainers have confirmed 24 of the 26 defects they have responded to, including ten previously unknown defects, nine of which are already fixed.
\end{abstract}


\maketitle
\fancyfoot{}

\section{Introduction}

Maintainers use Static Application Security Testing (SAST) reports to identify code that may require a security fix. Acting on a report requires understanding whether its suspected behavior can occur in the project. Accepting a false alarm consumes review effort, while dismissing a real defect leaves the program vulnerable. On the 200 most-starred C/C++ projects on GitHub, five SAST tools raise 45,440 alerts in the five vulnerability families that we study. Studies of static-analysis use show that the information available for investigating warnings affects whether developers act on them~\cite{johnson2013d,ayewah2009checklists,vassallo2018context,ami2024static,charoenwet2024secure}.

The decisive information often lies beyond the reported operation. Caller-selected constants may rule out a helper's null return, while initialization through another procedure may establish the state required by a later use. Production analyzers use abstractions and summaries to analyze large codebases~\cite{distefano2019scaling}, and their reports may omit these project-specific relationships. Large language models (LLMs) can interpret such code and project conventions once they see them, as illustrated by work on initialization warnings~\cite{llift2023} and graph-guided vulnerability detection~\cite{llmxcpg2025}. Existing LLM-based triage retrieves context for the model and adopts its verdict~\cite{wen2024llm4sa,llift2023}.

Using this ability for adjudication raises two needs. First, the model must see the decisive code, such as caller constants or helper effects, although the report names only the operation where investigation starts. Second, its verdict needs a basis that program analysis can check. Consider a warning that a struct member may be used uninitialized after a helper call. Whether the helper's initialization satisfies the later use is a semantic judgment for the model. Which object the helper receives is a program relation that analysis can check, and a justification that attributes another object's initialization to the reported member fails this check.

We present \textsc{VulnSAGE}, an adjudication system for SAST reports that connects LLM reasoning to program structure in both directions. Program structure determines what the model reads, and checks of program structure decide which of its interpretations count toward a verdict.

\noindent\textbf{We slice from the report along dependencies and the structure that explains them.}
Dependencies lead from the reported operation to the origins of its values. Deciding the report also requires the structure that gives these values meaning. Copy lengths may depend on table entries, while the destination capacity comes from a declaration outside that dependence chain. Helper initialization matters for a report through the binding of the caller's object to the helper's parameter. We introduce a \emph{Program Structure Evidence Graph (PSEG)} that represents dependencies together with the bindings and structure that explain them as labeled edges over shared program nodes. Our slicer traverses PSEG from the report's locations and gives the model an initial slice before it reasons. Because PSEG nodes keep their source locations, the model can cite what it reads.

\noindent\textbf{We decide reports from semantic facts whose program premises hold.}
We introduce \emph{CLAIM}, an intermediate representation for LLM report judgments. Each CLAIM fact carries the model's interpretation of how the reported condition arises or is ruled out, together with the program relations that this interpretation assumes as its structural premises. The model is responsible for the semantic part, while the structural premises denote PSEG relations that program analysis checks. Structural admission checks all cited premises and admits the fact when each holds and the cited relations reach the reported operation. The verdict follows from admitted facts, and missing support yields an uncertain outcome. Each decided report therefore rests on a fact whose cited program relations hold.

For statically supported true positives, \textsc{VulnSAGE} carries the trigger conditions into proof-of-concept (PoC) construction and execution. The resulting input and invocation let maintainers observe the reported behavior. In our confirmed libtiff case, a generated BigTIFF input reproduces the allocation/copy mismatch identified during adjudication under AddressSanitizer~\cite{serebryany2012asan}.

Across 187 labeled reports from four corpora covering five vulnerability families, \textsc{VulnSAGE} rejects 89 of 97 false alarms and retains 85 of 90 real defects, and it decides at least 96\% of the reports in each set. Report-centered pre-slicing improves all-report accuracy by 14.81-27.08 percentage points. Without CLAIM admission, false alarms promoted to positives rise from four to 14 on D2A-libtiff~\cite{d2a2021}, and missed real defects rise from one to three on SATE-V~\cite{delaitre2018satev}. Under a common PoC procedure, 25 positive verdicts of \textsc{VulnSAGE} reproduce at runtime, compared with three of the LLM-based adjudicator LLM4SA~\cite{wen2024llm4sa}. On the open-source alerts, the system reproduces 104 defects. Maintainers have confirmed 24 of the 26 defects they have responded to, including ten previously unknown defects, nine of which are already fixed.

This paper makes three contributions:
\begin{enumerate}
\item \textbf{Report-centered slicing over PSEG.} We introduce PSEG, a program representation whose edges cover dependencies and the relations that explain them, and a slicer that recovers report context from trace locations while retaining the source identity that CLAIM facts cite.
\item \textbf{CLAIM, an intermediate representation for LLM report judgments.} We define its syntax and give its structural premises a meaning over PSEG. Verdicts follow from admitted facts, which makes each decided verdict rest on a fact whose cited program relations hold.
\item \textbf{Adjudication with runtime reproduction and its evaluation.} We implement \textsc{VulnSAGE} to carry supported reports into PoC construction and execution. An evaluation on 187 labeled reports and 45,440 open-source alerts reproduces 104 defects and finds ten previously unknown defects confirmed by maintainers.
\end{enumerate}

\section{Problem Setting and Motivation}
\label{sec:challenges}

\subsection{Report Input and Outcomes}
\label{sec:report-evidence-boundary}

\textsc{VulnSAGE} adjudicates individual SAST reports against their matching source revisions. Each report supplies a vulnerability family and primary source location with diagnostic text. An optional trace supplies further locations for investigation, and the enclosing procedure refines the reported target. We target source-available C/C++ projects and five vulnerability families, namely buffer overflow, memory leak, invalid lifetime, null dereference and uninitialized state.

Adjudication determines whether project-specific information supports the reported defect or rules it out. False positives receive \texttt{FP}, while statically supported true positives receive \texttt{TP-S}. Missing support yields \texttt{uncertain}. An observation correlated with the reported operation can refine \texttt{TP-S} to \texttt{TP-D}, a dynamically reproduced true positive.

\subsection{Report-Specific Context}
\label{sec:report-context}

Figure~\ref{fig:motivation-cases} illustrates three false positives whose decisive information lies in different parts of the program. The snippets retain the relationships relevant to each decision.

\begin{figure*}[t]
\begin{lrbox}{\fpmpanelboxA}\begin{minipage}[t]{\fpmhalfinner}
\begin{lstlisting}[style=fpmpanel,emph={bits,poly,AV_CRC_16_ANSI}]
const Table *
av_crc_get_table(int alg) {
    ...
    if (av_crc_init(&table, ..., bits,
                    poly, ...) < 0)
        return NULL;
    return &table;
}
// AV_CRC_16_ANSI selects
// bits=16, poly=0x8005.
// init rejects bits outside [8,32]
// or poly >= (1LL << bits).
const Table *t =
    av_crc_get_table(AV_CRC_16_ANSI);
// Infer: possible NULL dereference
t->data[0];
\end{lstlisting}
\end{minipage}\end{lrbox}%
\begin{lrbox}{\fpmpanelboxB}\begin{minipage}[t]{\fpmhalfinner}
\begin{lstlisting}[style=fpmpanel,emph={err_levels,NGX_MAX_ERROR_STR}]
static ngx_str_t err_levels[] = {
    ngx_string("emerg"),  // len=5
    ngx_string("notice"), // len=6
    ...                   // all <= 6
};
ngx_str_t *v = &err_levels[level];
u_char buf[NGX_MAX_ERROR_STR]; // 2048

// Infer models v->len as [0,+inf).
ngx_cpymem(buf, v->data, v->len);
// Infer: possible buffer overrun
\end{lstlisting}
\end{minipage}\end{lrbox}%
\fpmpanelpair{(a) libav: Caller constants}{(b) nginx: Table values}
\medskip
\begin{lrbox}{\fpmpanelboxA}\begin{minipage}[t]{\fpmsplitinner}
\begin{lstlisting}[style=fpmpanel,emph={stNode,p}]
SMARTMC_LUDATA_NODE_S stNode;
smartmc_lucfg_LuDataInit(
    &stNode, kind, max_id, &param);
DTQ_AddTail(&stList, &(stNode.stNode));
// SAST: member may be uninitialized
\end{lstlisting}
\end{minipage}\end{lrbox}%
\begin{lrbox}{\fpmpanelboxB}\begin{minipage}[t]{\fpmsplitinner}
\begin{lstlisting}[style=fpmpanel,emph={stNode,p}]
static void smartmc_lucfg_LuDataInit(
    SMARTMC_LUDATA_NODE_S *p, ...) {
    DTQ_NodeInit(&(p->stNode));
    p->enIdentifier = kind;
    p->usInstanceID = max_id;
    ...
}
\end{lstlisting}
\end{minipage}\end{lrbox}%
\fpmpanelsplit{(c) Helper-mediated initialization: Caller (left) and helper (right)}
\caption{Three simplified false-positive reports requiring different context: caller constants in libav, table values in nginx, and helper-mediated initialization. The helper case links the caller's object to the member passed to an initialization routine. Highlighted identifiers carry the decisive values. Unrelated code is omitted.}
\label{fig:motivation-cases}
\Description{Three source examples show report-specific context. The libav call supplies constants satisfying the helper's input constraints. The nginx table limits copy lengths to six bytes for a 2048-byte buffer. The helper example, shown as caller and callee, passes the local object's member to an initialization routine before its reported use.}
\end{figure*}

\paragraph{Call-site reasoning in libav.}
Figure~\ref{fig:motivation-cases}(a) shows a possible null return when \texttt{av\_crc\_init} rejects its inputs. At the reported call site, \texttt{AV\_CRC\_16\_ANSI} selects \texttt{bits=16} and \texttt{poly=0x8005}. These values satisfy the helper's constraints, making its failure guards false. The decision depends on how the supplied arguments relate to the rejection conditions. Argument bindings associate the caller's constants with the callee's parameters, while control context supplies the conditions used to interpret the possible null return.

\paragraph{Project data in nginx.}
In Figure~\ref{fig:motivation-cases}(b), the copy length comes from an entry in \texttt{err\_levels}. Its longest string has six bytes, while the destination holds 2,048 bytes. Infer~\cite{distefano2019scaling} models the length as an interval with no finite upper bound. The decision requires identifying what determines the length and available capacity. Data dependence connects the length to its table entry, while the table and destination declarations supply the values for comparison.

\paragraph{Helper-mediated initialization.}
Figure~\ref{fig:motivation-cases}(c) shows a list-node member consumed after a helper call. The helper invokes \texttt{DTQ\_NodeInit} on \texttt{p->stNode}. The decision requires determining which caller member receives this initialization. The binding from \texttt{\&stNode} to \texttt{p} identifies the object, and member context identifies the part passed to the initializer. We use the helper throughout the design to explain slicing and CLAIM admission.

The three cases share a pattern. The report locates the operation, and program relations lead from it to the decisive code. Dependencies recover the origins of values, such as the table entry that supplies the nginx length. Deciding each report also requires the structure that gives these values meaning, such as the call bindings in libav and the helper case. The nginx capacity comes from the destination's declaration, outside the dependencies of the length. None of this code appears at the reported line, and a model that retrieves code on demand must first suspect that it matters. We therefore recover dependencies with their explanatory structure from the report before the model reasons. Targeted retrieval then follows questions that arise during interpretation. Removing this initial context lowers adjudication accuracy by 14.81 to 27.08 percentage points (Section~\ref{sec:eval-ablation}).

\subsection{Semantic Facts and Program Premises}
\label{sec:facts-premises}

The helper also shows how a semantic interpretation depends on program relations. The model judges that the helper's initialization supplies the state that the consumer requires. Program analysis cannot decide this judgment in general~\cite{rice1953classes}. The judgment holds for the report only if the helper receives the caller's object and passes the reported member to the initializer. These are program relations that analysis can check. Interpretations that attribute another object's initialization to the reported member can sound equally plausible, but their object binding fails. Accepting such interpretations can promote false alarms or dismiss real defects. Section~\ref{sec:eval-ablation} shows both effects when admission is bypassed.

CLAIM therefore attaches each semantic fact to the program relations it assumes. Structural admission checks these premises, and the verdict follows from facts whose premises hold. The premises cite program elements in the context recovered for the report, and checks of these relations decide whether a semantic interpretation counts toward the verdict.

\subsection{Static Support and Runtime Reproduction}

Maintainers also need to observe the behavior of supported TPs. Bug-report studies identify reproduction information as valuable to resolution~\cite{bettenburg2008good,breu2010information}. Static interpretation supplies trigger conditions that PoC construction encodes in an input and invocation. For the previously unknown libtiff vulnerability confirmed by upstream maintainers, adjudication traces wide metadata through a narrower allocation bound to a later copy, and a BigTIFF input with its configured invocation reproduces this behavior under AddressSanitizer (Section~\ref{sec:case-study-libtiff54}).

\subsection{Design Requirements}

\textbf{R1: Recover dependencies with the structure that explains them.}
Follow report locations to the program context needed for interpretation, retaining the source identity of each element.

\textbf{R2: Decide from facts whose program premises hold.}
Tie each semantic fact to the program relations it assumes, and compute the verdict from facts that pass structural admission.

\textbf{R3: Seek runtime reproduction for supported TPs.}
Carry static trigger conditions into PoC construction and execution to seek an observation correlated with the report.

\section{Design}
\label{sec:approach}

\subsection{Overview}
\label{sec:approach_overview}

\textsc{VulnSAGE} connects LLM reasoning to program structure in two directions. From program to model, a report-centered slicer recovers the code that the model needs to interpret the report. From model to program, CLAIM records each interpretation as a semantic fact that cites the program relations it assumes, and structural admission checks these relations before the fact can support a verdict. Both directions operate on PSEG, a program graph for report investigation whose nodes retain their source identity, that is, their file, line and code. The code that the model reads is therefore code that it can cite, and the relations it cites are relations that the checker can query.

\begin{figure*}[t!]
\centering
\includegraphics[width=\textwidth]{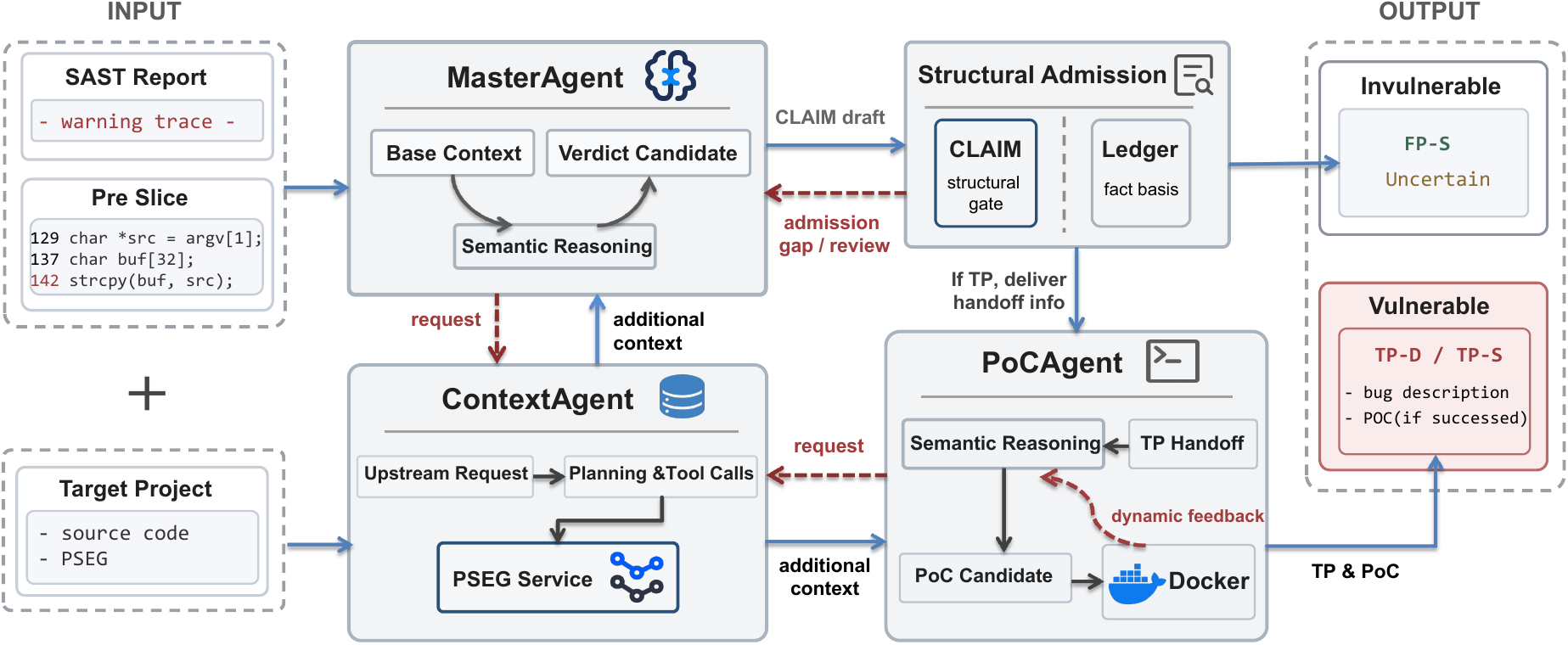}
\Description{A SAST report and its source revision enter the PSEG service, which resolves locations and builds the report-centered slice. The ContextAgent retrieves further context, and the MasterAgent writes CLAIM facts whose structural premises are checked before the ledger computes the verdict. Statically supported true positives proceed to PoC generation and execution, which can reproduce the reported behavior.}
\caption{\label{fig:vulnsage-pipeline}Workflow of \textsc{VulnSAGE}. A SAST report and its source revision enter the PSEG service, which builds the report-centered slice. The ContextAgent retrieves further context on request, and the MasterAgent writes CLAIM facts whose cited premises pass through admission before the ledger computes the verdict. Each \texttt{TP-S} enters PoC generation and execution, and a correlated runtime observation yields \texttt{TP-D}.}
\end{figure*}

Figure~\ref{fig:vulnsage-pipeline} shows the workflow, which we follow for the helper report of Figure~\ref{fig:motivation-cases}(c). The slicer starts at the reported member use and recovers the preceding helper call with the binding of \texttt{\&stNode} to parameter \texttt{p}. Targeted retrieval through the ContextAgent then fetches the initializer call on \texttt{p->stNode} (Section~\ref{sec:report-centered-slicing}). When the model stops requesting context, it proposes a \emph{candidate side}, here FP, together with a CLAIM. Its fact interprets the helper call as an initialization that rules out the warning and cites the binding and member relations (Section~\ref{sec:claim-grounding}). Admission checks these relations through PSEG and admits the fact, or returns feedback for citation repair. The ledger computes the verdict from admitted facts and returns \texttt{FP} (Section~\ref{sec:structural-admission}). A \texttt{TP-S} verdict proceeds to PoC construction and execution, where a correlated runtime observation yields \texttt{TP-D} (Section~\ref{sec:validation-outcomes}).

\subsection{Report-Centered Slicing}
\label{sec:report-centered-slicing}

Program slicing collects the statements that may affect a value at a program point~\cite{weiser1984slicing}. Report adjudication needs these dependencies together with \emph{explanatory relations}, which connect a dependence to the declaration, object or guard that determines its meaning for the report (Section~\ref{sec:report-context}). Each kind of relation answers a question about the suspected state, such as a possibly null value or an uninitialized member. Data dependence shows where a value comes from and how it is transformed. Control relations locate the guards or exits that govern the operation. Call bindings identify the caller object or value on which a callee acts. Declarations and type/member relations supply definitions such as integer widths or the member being accessed.

Our slicer differs from dependence slicing in three design choices. It follows explanatory relations in addition to dependencies, both represented as PSEG edges. It takes its criteria from the report and uses each trace location as a seed. It returns source excerpts whose program elements retain their identity, which lets CLAIM facts cite them.

\subsubsection{PSEG Structure and Construction}

Report investigation moves from a use to its definition, or from a caller's actual argument to a member accessed in the callee. PSEG represents each kind of relation listed above as a group of labeled edges over shared program nodes, including the parameter bindings that interprocedural slicing follows~\cite{horwitz1988interprocedural}. Storing all of them as edges lets a bounded traversal from report seeds recover dependencies together with their explanatory relations, while CLAIM relation queries inspect the graph that supplied the model's context. We define PSEG as
\begin{equation}
\begin{aligned}
G&=(V,E,\alpha),\\
E&=E_{\mathrm{syn}}\cup E_{\mathrm{cfg}}\cup E_{\mathrm{dep}}
  \cup E_{\mathrm{call}}\cup E_{\mathrm{sym}}\cup E_{\mathrm{ctx}}.
\end{aligned}
\label{eq:pseg}
\end{equation}
The shared node set $V$ contains program expressions and statements, with declarations and procedure boundaries. Formal parameters and actual arguments identify the ends of call bindings. Type declarations and members describe object structure. The attribute map $\alpha$ retains source locations and code, together with symbol/type information and argument indices. These attributes support both source reconstruction and endpoint resolution.

$E_{\mathrm{syn}}$ holds abstract syntax tree (AST) edges and $E_{\mathrm{cfg}}$ the control-flow graph (CFG), which together give the base program structure. The other groups hold the relations that slicing follows. $E_{\mathrm{dep}}$ contains reaching definitions and control dependence~\cite{ferrante1987pdg}. Resolved calls and argument positions determine $E_{\mathrm{call}}$. Declarations and type membership determine $E_{\mathrm{sym}}$. The group $E_{\mathrm{ctx}}$ links operations to their surrounding source structure. Distinct edge labels preserve the meaning of relations between shared endpoints.

Table~\ref{tab:pseg-construction} uses program-node roles to summarize construction. Roles may share a node type. The schema $X\to Y$ denotes a stored directed edge between matching nodes under the construction conditions below. A slash separates alternative roles, and $i$ denotes the matching formal/actual parameter position at a resolved call. The traversal policy in the next subsection determines how queries follow these edges.

\begin{table}[t]
\centering
\caption{PSEG construction rules as directed edge schemas. Arrows indicate stored edge directions.}
\label{tab:pseg-construction}
\begin{tabularx}{\linewidth}{@{}l >{\raggedright\arraybackslash}X@{}}
\toprule
Relation group & Directed edge schemas \\
\midrule
Dependence & Reaching-definition, control-dependence and condition edges. \\
\addlinespace[3pt]
Call boundary & $\mathrm{Method}\to\mathrm{CallSite}$,\quad
                $\mathrm{ReturnBoundary}\to\mathrm{CallSite}$ \\
\addlinespace[3pt]
Parameter binding & $\mathrm{FormalIn}_i\to\mathrm{Actual}_i$,\quad
                    $\mathrm{FormalOut}_i\to\mathrm{Actual}_i$ \\
\addlinespace[3pt]
Declaration & $\mathrm{Decl}\to\mathrm{IdentifierUse}$,\quad
              $\mathrm{GlobalDef}\to\mathrm{ExternDecl/Ref}$ \\
\addlinespace[3pt]
Type/member & $\mathrm{TypeDecl}\to\mathrm{Local/Member}$,\quad
              $\mathrm{ContainingType}\to\mathrm{Member}$ \\
\addlinespace[3pt]
Indirect call & $\mathrm{PointerBinding}\to\mathrm{IndirectCall}$ \\
\addlinespace[3pt]
Control context & $\mathrm{Condition}\to\mathrm{ControlStructure}$,\quad
                  $\mathrm{Case/Default}\to\mathrm{CFGNode}$ \\
\addlinespace[3pt]
Source structure & $\mathrm{Block}\to\mathrm{ASTDescendant}$,\quad
                   $\mathrm{OperatorArgument}\to\mathrm{Operator}$ \\
\bottomrule
\end{tabularx}
\end{table}

Construction starts from reaching-definition, control-dependence and condition edges. For a resolved call, we connect the callee node ($\mathrm{Method}$) and its method-return boundary node ($\mathrm{ReturnBoundary}$) to the calling expression ($\mathrm{CallSite}$). The callee's input/output formal-parameter nodes ($\mathrm{FormalIn}_i$, $\mathrm{FormalOut}_i$) connect to the actual argument expression ($\mathrm{Actual}_i$) at matching position $i$. These edges record parameter correspondence, leaving the call's effect to interpretation of the callee code. For local function-pointer assignments with an identified callee, the assigned pointer's declaration or identifier ($\mathrm{PointerBinding}$) connects to its indirect call site ($\mathrm{IndirectCall}$). We also add call-boundary and parameter links for that callee, allowing retrieval to enter its context.

Symbol links connect declarations ($\mathrm{Decl}$) to their identifier uses ($\mathrm{IdentifierUse}$). Available global definitions ($\mathrm{GlobalDef}$) connect to corresponding external declarations ($\mathrm{ExternDecl}$) or references ($\mathrm{Ref}$). Using type names, we connect each type declaration ($\mathrm{TypeDecl}$) to local-variable declarations ($\mathrm{Local}$) or member declarations ($\mathrm{Member}$) of that type. Member declarations also receive an edge from the type declaration that contains them ($\mathrm{ContainingType}$). These links recover both the member's type and its place in the enclosing object.

For control context, condition expressions ($\mathrm{Condition}$) connect to their associated constructs ($\mathrm{ControlStructure}$), with further control links through if/else nesting. From each switch-label node ($\mathrm{Case/Default}$), CFG traversal adds an edge to each visited node ($\mathrm{CFGNode}$), stopping expansion at a break or method-return boundary. Block nodes ($\mathrm{Block}$) connect to their AST descendants ($\mathrm{ASTDescendant}$), including nested blocks at which descent stops. Operand expressions ($\mathrm{OperatorArgument}$) connect to the operation-expression nodes that use them ($\mathrm{Operator}$), retaining the operands needed to interpret expressions.

\subsubsection{Criteria and Traversal}

We use the report's locations as slicing criteria to focus retrieval on the operation being adjudicated. Let $A_r$ contain the distinct locations in report $r$'s trace, or its primary location if no trace is supplied. Resolution maps these criteria to seed nodes $N_r\subseteq V$. A file and line may identify several expressions, refined by column information when available. We retain their candidate nodes and record resolution gaps separately.

Let $K$ be the enabled slicing relations, comprising dependence edges and the constructed links in Table~\ref{tab:pseg-construction}. AST and CFG relations remain available for construction and queries. Backward traversal recovers sources and supporting context through $\operatorname{Next}_{-}(v)$, the predecessors of $v$ along $K$. Forward traversal follows successors to recover uses, except for call-boundary and parameter edges. At a call site it visits the callee boundary. At an actual argument it visits the formal input parameter. Formal output parameters participate in backward retrieval. This policy lets a forward query enter the callee from a caller value despite the stored formal-to-actual orientation.

For direction $\delta$ and maximum hop depth $d$, let $\operatorname{Reach}^{\delta}_{K,d}(N_r)$ be the nodes visited from $N_r$ under this policy. With $\mu$ mapping nodes to source lines, the graph-selected portion of the initial context is
\begin{equation}
B_r=\mu\!\left(\operatorname{Reach}^{-}_{K,d}(N_r)\right).
\label{eq:report-slice}
\end{equation}
Depth limits the number of relation hops explored per request. The initial slice uses $d=1$ for trace seeds, because consecutive trace locations already follow the analyzer's path, and $d=3$ when only the primary location is available. Targeted queries can start from newly identified locations when the initial context leaves a question unresolved.

\begin{algorithm}[t]
\caption{Multi-seed slicing over PSEG.}
\label{alg:pseg-slice}
\begin{algorithmic}[1]
\REQUIRE PSEG $G$, criteria $A_r$, depth $d$, directions $\Delta$
\STATE $(N_r,\Gamma)\leftarrow\mathrm{ResolveLocations}(G,\mathrm{distinct}(A_r))$
\STATE $T\leftarrow\varnothing$
\FORALL{$\delta\in\Delta$}
  \STATE $S\leftarrow N_r$, FIFO queue $W\leftarrow[(v,0)\mid v\in N_r]$
  \WHILE{$W\ne\varnothing$}
    \STATE $(v,h)\leftarrow\mathrm{pop}(W)$
    \IF{$h<d$}
      \FORALL{$u\in\operatorname{Next}_{\delta}(v)\setminus S$}
        \STATE $S\leftarrow S\cup\{u\}$, $\mathrm{push}(W,(u,h+1))$
      \ENDFOR
    \ENDIF
  \ENDWHILE
  \STATE $T\leftarrow T\cup S$
\ENDFOR
\STATE \textbf{return} $(\mu(T)\cup\mathrm{SourceAtGaps}(A_r,\Gamma),\Gamma)$
\end{algorithmic}
\end{algorithm}

Algorithm~\ref{alg:pseg-slice} uses $\Delta=\{-\}$ initially and $\Delta=\{-,+\}$ when a data-flow request asks for the uses of a value. Its visited set merges overlapping dependencies among seeds and terminates cycles. Breadth-first order reaches a node at its smallest hop distance for that direction. Excluding location lookup and source rendering, traversal costs $O(|V_{\mathrm{seen}}|+|E_{\mathrm{scanned}}|)$ per direction and uses $O(|V_{\mathrm{seen}}|)$ working memory.

\subsubsection{Source Reconstruction and Targeted Retrieval}
\label{sec:static-seeding-retrieval}

The map $\mu$ groups selected nodes by source file and line. It removes duplicate locations before rendering excerpts in source order. Each result retains its criterion and direction with source references. Resolution gaps $\Gamma$ identify unresolved locations or ambiguous matches. Source text available at such locations remains context for the model, with its graph-resolution status preserved. The slice reports structural reachability, whose security meaning the model interprets for the report.

For the helper in Figure~\ref{fig:motivation-cases}(c), retrieval starts at the member consumed by \texttt{DTQ\_AddTail}. Local context exposes the preceding helper call. Its argument binding associates \texttt{\&stNode} with parameter \texttt{p}. A callee query retrieves the use of \texttt{p->stNode} in \texttt{DTQ\_NodeInit}, while declaration and type queries identify the member. The excerpts let the model interpret the helper's initialization effect. Their source references let the resulting CLAIM fact cite the corresponding bindings, which admission checks through relation queries to the PSEG service.

The ContextAgent extends this context in response to a request that identifies an unresolved question and a report-relevant location or symbol. Requests can resolve a location, fetch a source window, slice from a new location in either direction, follow callers and callees, fetch a symbol definition or search the project code. Each response returns the requested source or relations with references and query status. Requests can recover caller-selected constants in libav or table entries in nginx. Retrieval ends at a terminal model candidate, after ten request rounds, or when repeated requests add no report-relevant information. Required context that remains missing is handled by the ledger policy.

\subsection{CLAIM Intermediate Representation}
\label{sec:claim-grounding}

CLAIM is the intermediate representation between the model's semantic reasoning and program-relation checking, and it determines the basis of each verdict. Following the division in Section~\ref{sec:facts-premises}, a CLAIM fact pairs a semantic judgment with the structural premises that it assumes. The checker validates the form of semantic components against a \emph{family template}, which lists the roles that a fact for the report's vulnerability family must fill (Table~\ref{tab:family-fact-shapes}). The model is responsible for their truth. Structural premises denote PSEG queries, and the checker decides whether they hold. Three principles govern how facts reach a verdict.

\emph{Facts come first.} The interpretation determines which relations matter, and analysis cannot tell which relations decide a report before an interpretation selects them. The helper's FP fact rests on an object binding, while a null-dereference TP rests on an assignment and a data flow. The model therefore cites the premises of each fact, and checking follows the fact.

\emph{Facts are admitted as units.} Admission checks all premises that a fact cites. It requires each to hold and at least one to relate program elements beyond identifying or ordering them. Report anchoring further requires a cited relation to reach the reported location or the reported object, which keeps true but unrelated relations from supporting a fact.

\emph{The ledger decides.} The \emph{evidence ledger} records admitted facts and computes the verdict from them once \emph{retrieval gates} confirm that the reported location resolves and its local source was retrieved. Missing support yields \texttt{uncertain}.

\subsubsection{Syntax and Fact Fields}
\label{sec:claim-syntax}

Equation~\ref{eq:claim-syntax} gives the abstract syntax of CLAIM.
\begin{equation}
\begin{array}{@{}r@{\;}c@{\;}l@{\qquad}l@{}}
C & ::= & (\mathcal{F},\mathcal{Q},\mathcal{I}) & \text{CLAIM}\\
f & ::= & \langle s,h,k,z,o,\pi,c,e,t,\mathcal{Q}_f,\mathcal{I}_f\rangle & \text{fact}\\
s & ::= & \mathrm{TP}\mid\mathrm{FP} & \text{side}\\
k & ::= & \textsf{source-to-sink}\mid\textsf{closure-to-sink} & \text{fact kind}\\
  & \mid & \textsf{closure-to-boundary}\mid\textsf{disjointness} & \\
c & ::= & \textsf{none}\mid\textsf{guard}\mid\textsf{early-exit}\mid\textsf{rewrite}\mid\textsf{initialize}\mid\textsf{rebind}\mid\textsf{release} & \text{closure}\\
  & \mid & \textsf{transfer}\mid\textsf{preserve}\mid\textsf{revalidate}\mid\textsf{replace}\mid\textsf{disjoint}\mid\textsf{boundary} & \\
q & ::= & \{a_1,\ldots,a_n\} & \text{premise bundle}\\
a & ::= & R_{\theta}(x,y)\mid R_{\theta}(x) & \text{atom (Table~\ref{tab:claim-relations})}\\
x,y & ::= & \langle \mathit{loc},\mathit{fn}^{?},\mathit{path}^{?},\mathit{role}\rangle & \text{source endpoint}
\end{array}
\label{eq:claim-syntax}
\end{equation}
CLAIM $C$ groups model-authored facts $\mathcal{F}$ with the premise bundles $\mathcal{Q}$ and source references $\mathcal{I}$ that they cite. Each fact selects its cited bundles $\mathcal{Q}_f\subseteq\mathcal{Q}$ and references $\mathcal{I}_f\subseteq\mathcal{I}$ through identifiers. Source and retrieval references locate the excerpts used in interpretation and the program elements referenced by atoms. Table~\ref{tab:helper-claim} defines the fact fields and instantiates them for the helper. Fields $s$, $h$, $k$, $z$, $o$ and $t$ state what the fact asserts. Fields $\pi$, $c$ and $e$ explain how it addresses the report, while $\mathcal{Q}_f$ and $\mathcal{I}_f$ identify what it rests on. The carrier $\pi$ lists the relation kinds that associate the relevant state with the reported operation, and the cited relations themselves are stored in $\mathcal{Q}_f$. Fields $z$ and $o$ range over finite vocabularies of source-state kinds, such as null or uninitialized, and sink kinds, such as dereference or copy, and $h$ names one of the family templates.

\begin{table}[t]
\centering
\caption{Fact fields and an illustrative helper CLAIM. The premise bundles are defined in the text.}
\label{tab:helper-claim}
\begin{tabularx}{\linewidth}{@{}p{0.12\linewidth}p{0.30\linewidth}X@{}}
\toprule
Field & Role & Helper instance \\
\midrule
$s,h$ & Side and family template & FP, initialization/rebinding/disjointness template. \\
$k$ & Fact kind & Closure-to-sink assertion. \\
$z,o$ & Source-state and sink kinds & Uninitialized storage, use. \\
$\pi$ & Carrier relationships & Call-argument binding and declaration/reference. \\
$c$ & FP closure operation & Initialization by the helper. \\
$e$ & Semantic explanation & The helper initializes the list-node member before the consumer uses it. \\
$t$ & Reported target & Use of the caller's list-node member by the consumer. \\
$\mathcal{Q}_f$ & Cited premise bundles & $\{q_{\mathrm{bind}},q_{\mathrm{member}}\}$. \\
$\mathcal{I}_f$ & Source/retrieval references & References to the caller and helper excerpts in Figure~\ref{fig:motivation-cases}(c). \\
\bottomrule
\end{tabularx}
\end{table}

The side $s$ selects the interpretation that the fact supports, and the family template $h$ constrains its roles (Table~\ref{tab:family-fact-shapes}). TP source-to-sink facts use $c=\mathrm{none}$ and explain how a state carrier reaches the reported operation. FP facts identify an intervening operation or disjointness that rules out the reported behavior. Their closure $c$ names the mechanism, with $e$ explaining how it addresses the warning.

\begin{table*}[t]
\centering
\caption{\label{tab:family-fact-shapes}Semantic roles of the ten family templates, one per family and side. The model supplies their interpretation, with cited relations checked during admission.}
\begin{tabularx}{\textwidth}{@{}p{0.17\textwidth}XX@{}}
\toprule
\textbf{Vulnerability family} & \textbf{TP interpretation} & \textbf{FP interpretation} \\
\midrule
Buffer overflow & A size or bounds carrier reaches the reported access with insufficient capacity. & A guard or rewrite rules out the mismatch, or the carrier is disjoint from the access. \\
Memory leak & An acquired resource reaches a boundary where ownership would be lost. & The resource is released, or its ownership is transferred or preserved. \\
Invalid lifetime & An invalidated resource reaches a later use. & Revalidation or replacement restores valid state. Preservation or disjointness separates the use from invalid state. \\
Null dereference & A null-producing state reaches the reported null-sensitive operation. & A guard or early exit excludes the null state, or rebinding supplies a non-null object. \\
Uninitialized state & Uninitialized storage reaches the reported read or use. & Initialization or rebinding supplies the state, or the use concerns disjoint storage. \\
\bottomrule
\end{tabularx}
\end{table*}

A family template is a tuple $h=\langle \phi_h, s_h, K_h, Z_h, O_h, C_h\rangle$ of a vulnerability family, a side and the fact kinds, source-state kinds, sink kinds and closures that it permits. Each of the five families has one TP and one FP template, which gives ten templates (Table~\ref{tab:family-fact-shapes}). They are defined per family and side, independently of projects and benchmarks. Template matching checks the roles of a fact against its template, where $\phi(r)$ is the family of report $r$:
\begin{equation}
\operatorname{MatchTemplate}(f,r)\Longleftrightarrow \phi_h=\phi(r)\land s=s_h\land k\in K_h\land z\in Z_h\land o\in O_h\land c\in C_h.
\label{eq:match-template}
\end{equation}
Templates organize the vulnerability-specific interpretation, while the shared relation vocabulary specifies the program connections to check.

\subsubsection{Structural Premises}

Each premise bundle $q\in\mathcal{Q}$ contains a set $\operatorname{Atoms}(q)$ of atoms that together establish one program connection of the fact, such as the member passed to the initializer. Atoms are neutral. They state program relations without security meaning, which the explanation $e$ assigns. An atom is $a=R_{\theta}(x,y)$, where $x$ and $y$ are source endpoints and $\theta$ supplies parameters such as a callee or argument position. We identify an endpoint by its file and line, optionally refined by function and symbol/access-path information. Each endpoint's role records its purpose in the fact, such as the reported target or a carrier, and report anchoring uses it. Source endpoints let the model cite retrieved code, while the checker resolves local PSEG nodes for the query.

\begin{table}[t]
\centering
\caption{CLAIM's neutral relation vocabulary. Short relation names denote the indicated queries. The last column lists the PSEG relation groups of Equation~\ref{eq:pseg} or the attributes that each query inspects. $^{\dagger}$ marks auxiliary relations, which cannot admit a fact alone.}
\label{tab:claim-relations}
\begin{tabularx}{\linewidth}{@{}p{0.27\linewidth}X p{0.2\linewidth}@{}}
\toprule
Relation & Query over resolved endpoints & PSEG basis \\
\midrule
$\mathrm{DataReaches}(x,y)$ & A data relation connects carrier $x$ to operation $y$ under the configured traversal. & $E_{\mathrm{dep}}$, $E_{\mathrm{call}}$, $E_{\mathrm{sym}}$, $E_{\mathrm{ctx}}$ \\
$\mathrm{DeclRef}(x,y)$ & Reference $y$ is associated with declaration $x$. & $E_{\mathrm{sym}}$ \\
$\mathrm{Assign}(x,y)$ & An assignment binds the value at $x$ to the target at $y$. & $E_{\mathrm{syn}}$ \\
$\mathrm{Calls}(x,m)$ & Call $x$ resolves to method $m$. & $E_{\mathrm{syn}}$, callee in $\alpha$ \\
$\mathrm{CallArg}_{m,i}(x,y)$ & The expression at $x$ supplies argument position $i$ of $m$, corresponding to formal or argument-role endpoint $y$. & $E_{\mathrm{syn}}$, positions in $\alpha$ \\
$\mathrm{CallRet}(x,y)$ & The call result at $x$ is bound to the receiving value at $y$. & $E_{\mathrm{syn}}$ \\
$\mathrm{Control}(x,y)$, $\mathrm{Dominates}(x,y)$ & Control dependence or dominance relates $x$ to $y$. & $E_{\mathrm{dep}}$, $E_{\mathrm{cfg}}$, $E_{\mathrm{syn}}$ \\
$\mathrm{Exists}(x)^{\dagger}$, $\mathrm{SameSymbol}(x,y)^{\dagger}$, $\mathrm{SameObject}(x,y)^{\dagger}$ & Endpoint existence, symbol identity or matching object shape. & $\alpha$, endpoint fields \\
$\mathrm{MayPrecede}(x,y)^{\dagger}$ & Local source order allows $x$ to precede $y$. & Endpoint locations \\
\bottomrule
\end{tabularx}
\end{table}

Table~\ref{tab:claim-relations} defines the checking vocabulary. Each predicate denotes a query through the PSEG service that may inspect a graph pattern or traverse several edges. Data-reaching and declaration queries traverse the dependence, call and declaration edges that the slicer follows. Call and assignment queries match syntax patterns with argument positions, while identity and order relations compare endpoint fields. The relations that a fact cites therefore belong to the graph that supplied the model's context. Relation checks use the endpoint roles and parameters appropriate to the query. For example, a call-argument check uses the callee identity and argument index to distinguish multiple calls or parameters. Control and data relations can justify the program connections in a fact. The model supplies their security interpretation, including whether initialization is adequate or a capacity is sufficient.

\paragraph{Helper instance.}
Let $a$ identify the caller's \texttt{\&stNode} argument and $p$ the helper's formal parameter. Let $m$ identify the initializer's actual argument \texttt{\&(p->stNode)}, with $u$ denoting the reference to $p$ within that expression. Let $\nu$ denote the initializer's first argument role. With zero-based argument positions, the fact in Table~\ref{tab:helper-claim} cites
\begin{equation}
\begin{aligned}
\operatorname{Atoms}(q_{\mathrm{bind}})&=\{\mathrm{CallArg}_{\mathrm{helper},0}(a,p)\},\\
\operatorname{Atoms}(q_{\mathrm{member}})&=\{\mathrm{DeclRef}(p,u),\ \mathrm{CallArg}_{\mathrm{init},0}(m,\nu)\}.
\end{aligned}
\label{eq:helper-premises}
\end{equation}
Here $\mathrm{helper}$ denotes \texttt{smartmc\_lucfg\_LuDataInit}, with $\mathrm{init}$ denoting \texttt{DTQ\_NodeInit}. The caller reference locates $a$ and target $t$. The helper reference locates the parameter and member expressions, including the initializer call for $\nu$. Access-path information specifies the member received by the initializer.

For the TP in Figure~\ref{fig:claim-fact-first}, the record selects the null-dereference TP template with a source-to-sink kind. It identifies a null-producing state and a dereference sink. Assignment and data-reaching relations support its carrier, with $c=\mathrm{none}$. Both examples instantiate the fact form of Equation~\ref{eq:claim-syntax} with different semantic roles and premises.

\subsection{Admission Semantics and Verdict}
\label{sec:structural-admission}
\label{sec:evidence-bounded-judgment}

\begin{figure*}[t]
\centering
\includegraphics[width=\textwidth]{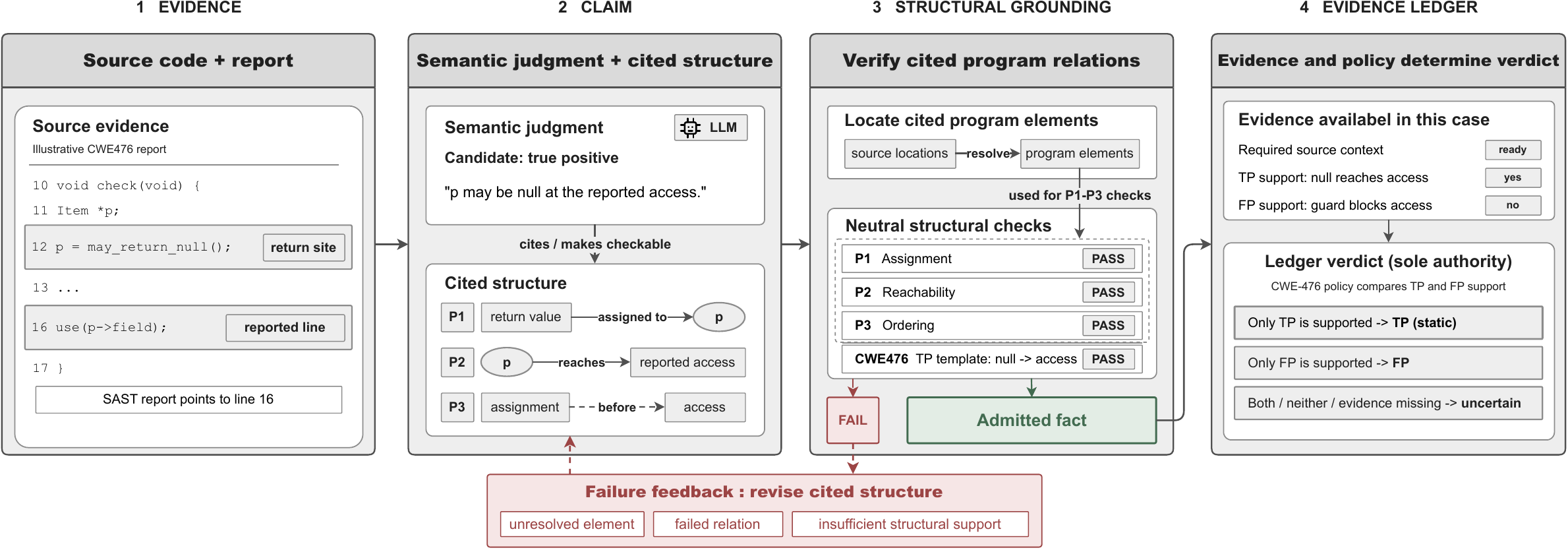}
\Description{A model-authored semantic fact cites source endpoints and neutral program relations. Admission checks these premises over PSEG, admitted facts enter the evidence ledger, and the ledger maps TP and FP support to the static verdict.}
\caption{\label{fig:claim-fact-first}Fact-first CLAIM admission for an illustrative null-dereference report. The model writes a semantic judgment and cites neutral program relations, which admission checks over PSEG. An admitted fact enters the evidence ledger, which maps TP and FP support to \texttt{FP}, \texttt{uncertain} or \texttt{TP-S}.}
\end{figure*}

This section gives CLAIM its meaning. Admission resolves the cited endpoints, interprets each premise over PSEG and then applies report-specific conditions. Facts that pass become \emph{admitted facts} in the ledger. Figure~\ref{fig:claim-fact-first} illustrates these stages for a null-dereference TP, and the helper instance below shows how they support an FP decision.

\subsubsection{Premise Interpretation}

For endpoint $x$, resolution returns candidate nodes $\operatorname{Resolve}_G(x)$ and a status. Source location restricts the search, while the function or symbol/access path distinguishes expressions at that location. Member expressions preserve their base and field information. For a call-argument role, the callee and argument index identify the corresponding source expression or formal parameter. When several nodes match, the relation check considers those that satisfy the endpoint constraints. A check that cannot resolve its endpoints returns unresolved.

Given the resolved endpoints, the checker evaluates $R_{\theta}$ with the query in Table~\ref{tab:claim-relations}. The result records supporting program references when available, or a reason the relation could not be established.

The interpretation $[\![a]\!]_G=\rho_G(a)$ of atom $a$ over $G$ is its relation status. Supported means the query establishes the cited relation. Refuted records a negative relation result for resolved endpoints, while unavailable covers unsupported queries or missing graph information. Unresolved covers endpoint ambiguity or an exhausted search. Identity and source-order relations are auxiliary. They can pass and help locate program elements, but they do not relate the elements through program structure. These results concern the program basis of the fact, whose security interpretation is model-authored.

\subsubsection{Admission Judgment}

We preserve the association between a fact and all its premises during checking. For fact $f$, define
\begin{equation}
\mathcal{A}_f=\bigcup_{q\in\mathcal{Q}_f}\operatorname{Atoms}(q).
\label{eq:fact-atoms}
\end{equation}
This set is determined by citations, independently of their check results. Let $\operatorname{Pass}_G(a)$ hold when $\rho_G(a)$ is supported. Let $\operatorname{Eligible}(a)$ hold when $a$ is not auxiliary, such as an argument binding or a data or control relation. Then
\begin{equation}
\begin{aligned}
\operatorname{NeutralSupport}_G(\mathcal{A}_f)\Longleftrightarrow{}&
\bigl(\forall a\in\mathcal{A}_f:\operatorname{Pass}_G(a)\bigr)\\
&\land\bigl(\exists a\in\mathcal{A}_f:\operatorname{Eligible}(a)\bigr).
\end{aligned}
\label{eq:neutral-support}
\end{equation}
The universal condition preserves the fact's complete cited basis. The existential condition requires a relation beyond locating, identifying or ordering program elements. Identity and source-order atoms can pass, but they cannot admit a fact alone. Any refuted, unavailable or unresolved cited atom blocks admission.

A fact is well formed when its fields take values from Equation~\ref{eq:claim-syntax} and its identifiers refer to bundles and references of its CLAIM. For a well-formed fact, report $r$ and the model's candidate side $j$, admission is the judgment
\begin{equation}
\frac{\operatorname{MatchTemplate}(f,r)\qquad s(f)=j\qquad \operatorname{Anchor}_{r}(\mathcal{Q}_f)\qquad \operatorname{NeutralSupport}_{G}(\mathcal{A}_f)}{\operatorname{Admit}_{G,r,j}(f)}
\label{eq:claim-admission}
\end{equation}
Template matching follows Equation~\ref{eq:match-template}. The side condition keeps admission attached to the current candidate. The target $t$ of each fact is the primary location of report $r$. Report anchoring $\operatorname{Anchor}_r(\mathcal{Q}_f)$ requires some cited atom to have an endpoint on the file and line of $t$, or at the reported object, that is, in the file and function of $t$ with its symbol or access path. In the helper case, the actual argument at $a$ carries the caller's object whose member appears at $t$. This link makes the checked helper binding relevant to the report.

\begin{algorithm}[t]
\caption{CLAIM admission with complete premise checking.}
\label{alg:claim-admission}
\begin{algorithmic}[1]
\REQUIRE CLAIM $C$, PSEG $G$, report $r$, candidate side $j$
\STATE $\Phi\leftarrow$ field and reference validation failures in $\mathcal{F}$
\FORALL{well-formed $f\in\mathcal{F}$}
  \FORALL{$a\in\mathcal{A}_f$}
    \STATE Resolve the endpoints of $a$, evaluate $[\![a]\!]_G$ and record it in $\Phi$
  \ENDFOR
  \STATE Record the admission conditions that $f$ fails in $\Phi$
\ENDFOR
\STATE $\mathcal{A}\leftarrow\{f\in\mathcal{F}\mid f\text{ is well formed}\land\operatorname{Admit}_{G,r,j}(f)\}$
\STATE \textbf{return} $(\mathcal{A},\Phi)$
\end{algorithmic}
\end{algorithm}

Algorithm~\ref{alg:claim-admission} evaluates all cited atoms of each well-formed fact, including atoms after a failed one, which gives repair feedback for each failure.

\subsubsection{Repair and Verdict}

For the helper instance, suppose both call bindings and the declaration-reference atom in Equation~\ref{eq:helper-premises} pass. With matching family roles and report anchoring, the fact enters the ledger as FP support. For an illustrative erroneous-object variant, suppose the cited actual argument refers to another object and the binding to the reported carrier is refuted. That atom remains in $\mathcal{A}_f$, causing the fact to fail admission even if its other atoms pass.

If a candidate contains facts but none passes, the checker returns feedback identifying failed endpoints or relations. In one repair round, the model may repair the citations and source references while preserving the candidate side. The repaired CLAIM undergoes the full gate again. The failed fact supplies no verdict support, and the outcome depends on other admitted facts and required context.

The ledger combines fact support with two mandatory retrieval gates that hold for all families. The reported location must resolve to PSEG nodes, and the source window around it must be retrieved. Let $M_r$ state that these gates hold. For the report's admitted facts $\mathcal{A}_r$ and side $x\in\{\mathrm{TP},\mathrm{FP}\}$, the support predicate $P_x(r)$ holds when some $f\in\mathcal{A}_r$ satisfies $s(f)=x$ and $\operatorname{MatchTemplate}(f,r)$. Because admission requires each fact to support the candidate side, admitted facts never support both sides. The static verdict follows
\begin{equation}
v_s(r)=
\begin{cases}
\texttt{TP-S}, & M_r\land P_{\mathrm{TP}}(r),\\
\texttt{FP}, & M_r\land P_{\mathrm{FP}}(r),\\
\texttt{uncertain}, & \text{otherwise.}
\end{cases}
\label{eq:ledger-policy}
\end{equation}
The admitted helper fact supports FP when the mandatory context is present. If its erroneous-object variant is the sole proposed fact and repair fails, the ledger lacks admitted support and returns uncertain.

\paragraph{Guarantees.} Relative to the relation checker, these definitions make each decided verdict rest on a semantic fact whose cited atoms all pass in PSEG (Equation~\ref{eq:neutral-support}). A \texttt{TP-S} or \texttt{FP} verdict requires the retrieval gates and an admitted fact for its side (Equation~\ref{eq:ledger-policy}). Admission and the static verdict are functions of the CLAIM, PSEG, the report and its retrieved context. The model therefore influences a verdict only through the facts, citations and retrieval requests that it writes.

\subsection{Runtime Reproduction}
\label{sec:validation-outcomes}

We carry each \texttt{TP-S} into a runtime test of its reported behavior. The handoff retains the report target and the admitted static basis, with its trigger conditions and source context. The PoCAgent uses this information to construct an input or harness, specifying an invocation and expected observation. Requests for missing input-format or execution information go through targeted retrieval. The runner executes the test against the selected revision and configuration. Section~\ref{sec:case-study-libtiff54} follows this handoff for the confirmed libtiff vulnerability.

Let $D$ denote the runtime result. The predicate $\operatorname{Correlated}(D,r)$ holds when a sanitizer report or crash in $D$ has a stack frame at the reported operation, on its file and line or in its function, while the project build of the selected revision runs the invocation that the PoC specifies. It refines the verdict as follows:
\begin{equation}
v(r,D)=
\begin{cases}
\texttt{TP-D}, & v_s(r)=\texttt{TP-S}\land\operatorname{Correlated}(D,r),\\
v_s(r), & \text{otherwise.}
\end{cases}
\label{eq:runtime-refinement}
\end{equation}
A correlated observation strengthens the admitted static TP basis with a reproduction. A clean run or unsuccessful execution preserves the static verdict because the attempted input may not reach the required state.

\section{Implementation}
\label{sec:implementation}

We implement \textsc{VulnSAGE} in Python using LangGraph~\cite{langgraph2024}. Report normalization extracts the vulnerability family and source location together with the diagnostic text and optional trace. Analysis operates on the source revision associated with the report. Reports from a multi-finding analyzer output are processed independently.

\paragraph{Program queries.}
We build PSEG on the code property graph (CPG) of Joern~\cite{yamaguchi2014cpg} for each source revision, reusing its C/C++ frontend and the program-structure and dependence edges that it computes. The CPG leaves the other PSEG relations implicit. Argument correspondence follows from matching positions, and the type of a member follows from its type name. The statements governed by a switch label require a walk along control flow. Overlay passes add these relations as the edges in Table~\ref{tab:pseg-construction}. The PSEG service runs location queries, traversal and graph relation checks on the Joern server, returning source locations for reconstruction. Source and project-search operations supplement graph queries. Responses retain references and resolution status for CLAIM construction. The slice in Algorithm~\ref{alg:report-level-detection} uses Algorithm~\ref{alg:pseg-slice}.

\paragraph{Workflow coordination.}
\label{sec:agent-orchestration}
The Supervisor routes requests between the MasterAgent and ContextAgent. The MasterAgent interprets reports and constructs CLAIMs, while the ContextAgent retrieves source and program context through the PSEG service. The MasterAgent prompt supplies the CLAIM schema with its family templates and relation vocabulary, and the model returns each candidate as a JSON object. For \texttt{TP-S}, the PoCAgent constructs a test and can request execution context through the ContextAgent. Algorithm~\ref{alg:report-level-detection} coordinates these stages. Its configuration $E$ comprises family fact policy and mandatory retrieval gates.

\begin{algorithm}[t]
\caption{Report adjudication and runtime reproduction.}
\label{alg:report-level-detection}
\begin{algorithmic}[1]
\REQUIRE report $r$, matching source tree $P$, PSEG $G$
\STATE $E\leftarrow$ family policy and retrieval gates for $r$
\STATE $L\leftarrow$ empty evidence ledger
\STATE $B\leftarrow\mathrm{PSEGSlice}(P,G,r)$
\REPEAT
  \STATE $(j,C,q)\leftarrow\mathrm{MasterAgent}(r,B)$
  \IF{$q$ requests retrievable report context}
    \STATE $B\leftarrow B\cup\mathrm{ContextAgent}(q,P,G)$
  \ENDIF
\UNTIL{$j$ is terminal or retrieval stops}
\IF{$j\in\{\mathrm{TP},\mathrm{FP}\}$}
  \STATE $(\mathcal{A},\phi)\leftarrow\mathrm{Admit}(C,G,r,j)$ using Algorithm~\ref{alg:claim-admission}
  \IF{$\mathcal{A}=\varnothing$ and $\phi$ permits structural repair}
    \STATE $C\leftarrow\mathrm{MasterAgentRepair}(C,\phi,j)$
    \STATE $(\mathcal{A},\phi')\leftarrow\mathrm{Admit}(C,G,r,j)$
  \ENDIF
  \STATE $L\leftarrow L\cup\mathcal{A}$
\ENDIF
\STATE $v_s\leftarrow\mathrm{LedgerPolicy}(E,L,B)$
\IF{$v_s=\texttt{TP-S}$}
  \STATE $D\leftarrow\mathrm{PoCAgent}(r,B,L)$ with targeted retrieval
  \IF{$\operatorname{Correlated}(D,r)$}
    \STATE \textbf{return} \texttt{TP-D}
  \ENDIF
\ENDIF
\STATE \textbf{return} $v_s$
\end{algorithmic}
\end{algorithm}

\paragraph{CLAIM processing.}
Schema validation checks fact fields and reference syntax before Algorithm~\ref{alg:claim-admission} applies the admission judgment. The ledger then applies Equation~\ref{eq:ledger-policy} to admitted facts and retrieved context.

\paragraph{Execution.}
The runner binds a test to its project revision and invocation, and its telemetry determines the refinement in Equation~\ref{eq:runtime-refinement}.

\section{Evaluation}
\label{sec:evaluation}

We address four research questions.

\textbf{RQ1: Adjudication quality.}
How well does \textsc{VulnSAGE} reject false alarms while retaining real defects, and how often does it decide?

\textbf{RQ2: Report-centered slicing and CLAIM admission.}
How do report-centered pre-slicing and CLAIM admission affect adjudication errors?

\textbf{RQ3: Reproducible verdicts.}
How do \textsc{VulnSAGE} and an LLM-based adjudicator compare in rejecting false alarms and in turning positive verdicts into runtime reproductions?

\textbf{RQ4: Open-source validation.}
How many defects does \textsc{VulnSAGE} reproduce among the alerts that SAST tools raise on popular open-source projects, and do maintainers confirm them?

\subsection{Experimental Setup}
\label{sec:eval-setup}

\paragraph{Datasets.}
Table~\ref{tab:eval-datasets} lists the evaluation sets, whose unit is a SAST report with its matching source revision. Juliet 1.3 provides C/C++ test cases with known vulnerable and non-vulnerable semantics~\cite{black2018juliet}. We run Infer~\cite{distefano2019scaling} on these cases and sample 100 reports, stratified by CWE and reference label with a fixed seed. The test-case semantics serve as reference labels. The real-world sets sample the supported families, and each reference label holds for the matching source revision.

D2A~\cite{d2a2021} and SATE-V~\cite{delaitre2018satev} contribute reports from real projects. Our D2A set contains 48 Infer reports from libtiff, which we call D2A-libtiff. \citet{le2024d2anoise} report a noise level above 50\% in the automatically labeled vulnerabilities of D2A. We therefore re-adjudicate the 48 reports by tracing the reported operation and its program context against the source revision and, where available, the upstream fix. Three annotators label each report independently before the experiments and resolve disagreements by discussion. Six labels change, five from FP to TP and one from TP to FP. The 27 SATE-V reports come from Cppcheck~\cite{cppcheck} and carry the labels of the manual warning analysis in SATE~V.

ARVO~\cite{arvo2026} provides reproducible vulnerabilities with crashing inputs and developer fixes. Because its reports all match real vulnerabilities, it measures how many such reports \textsc{VulnSAGE} retains, as opposed to how many false alarms it rejects. Several Infer reports may match one vulnerability, and its sanitizer trace and patch expose different program endpoints. We therefore keep tasks with recoverable vulnerable and fixed sources, sanitizer traces, build context and normalized Infer reports, and align reports with vulnerabilities by hand. We compare the dangerous operation, object and value carrier, trigger, call trace and repair of each candidate, and a candidate enters the Gold set when these elements match the vulnerability. We keep the report closest to the failure endpoint of each vulnerability, which yields 12 findings from five projects.

Juliet covers all five vulnerability families. D2A-libtiff covers buffer overflow, null dereference and uninitialized state, while SATE-V covers all families except buffer overflow.

\begin{table}[t]
\centering
\caption{Evaluation sets and their report-generating analyzers. D2A-libtiff labels come from manual re-adjudication.}
\label{tab:eval-datasets}
\small
\begin{tabular}{llrrl}
\toprule
Dataset & Analyzer & Reports & Reference TP/FP & Used for \\
\midrule
Juliet 1.3 & Infer & 100 & 50/50 & RQ1, RQ3 \\
D2A-libtiff & Infer & 48 & 5/43 & RQ1-RQ3 \\
SATE-V & Cppcheck & 27 & 23/4 & RQ1-RQ3 \\
ARVO Gold & Infer & 12 & 12/0 & RQ1, RQ3 \\
Open-source scan & Five SAST tools & 45,440 & n/a & RQ4 \\
\bottomrule
\end{tabular}
\end{table}

\paragraph{Configuration.}
\textsc{VulnSAGE} and the LLM4SA baseline of RQ3 use GPT-5.5~\cite{openai2026gpt55} with an identical configuration. PoC construction and execution run in an x86 Linux container.

\paragraph{Metrics.}
For each set, we count the false alarms that \textsc{VulnSAGE} rejects, the real defects that it retains as positives and its \texttt{uncertain} verdicts. Decision coverage is the share of reports that receive a decided verdict~\cite{geifman2017selective}. All-report accuracy counts \texttt{uncertain} as an error. For the real-world sets, we also count the retained defects that reproduce at runtime and receive \texttt{TP-D}. Juliet cases need no PoC execution.

\subsection{RQ1: Adjudication Quality}
\label{sec:eval-effectiveness}

\begin{table}[t]
\centering
\caption{Adjudication quality on the labeled sets. Parentheses give the retained real defects that reproduce at runtime.}
\label{tab:full-results}
\small
\setlength{\tabcolsep}{4pt}
\begin{tabular}{lrrrrrr}
\toprule
& & False alarms & Real defects & & & \\
Dataset & Reports & rejected & retained & \texttt{uncertain} & Coverage & Accuracy \\
\midrule
Juliet & 100 & 47/50 & 50/50 & 1 & 99.00\% & 97.00\% \\
D2A-libtiff & 48 & 38/43 & 4/5 (3) & 1 & 97.92\% & 87.50\% \\
SATE-V & 27 & 4/4 & 21/23 (16) & 1 & 96.30\% & 92.59\% \\
ARVO Gold & 12 & n/a & 10/12 (6) & 0 & 100.00\% & 83.33\% \\
\midrule
Total & 187 & 89/97 & 85/90 (25) & 3 & & \\
\bottomrule
\end{tabular}
\end{table}

Table~\ref{tab:full-results} shows that \textsc{VulnSAGE} rejects 89 of the 97 reference false alarms and retains 85 of the 90 real defects. It decides at least 96\% of the reports in each set. On D2A-libtiff, where 43 of the 48 reports are false alarms, it rejects 38 of them and retains four of the five real defects. On the three real-world sets, 25 of the 35 retained defects also reproduce at runtime.

An \texttt{uncertain} verdict means that the admitted facts and the retrieved context support neither side. For example, retrieval did not recover the upstream index constraints of a Juliet CWE122 false alarm, and the ledger returned \texttt{uncertain} without a supported FP fact.

\takeaway{1}{\textsc{VulnSAGE} rejects most false alarms while keeping real defects, and it rarely abstains.}

\subsection{RQ2: Report-Centered Slicing and CLAIM Admission}
\label{sec:eval-ablation}

We compare the complete configuration with two ablations on D2A-libtiff and SATE-V, whose labels come from manual adjudication. The first removes report-centered pre-slicing. The second keeps the structured fact output but bypasses structural admission. Both keep on-demand retrieval and model generation. We omit PoC execution to isolate static adjudication. Table~\ref{tab:ablation} gives the confusion counts of each configuration.

\begin{table}[t]
\centering
\caption{Complete and ablated configurations on D2A-libtiff (48 reports) and SATE-V (27 reports). FP counts false alarms promoted to positives, FN counts real defects rejected as FP, and U counts \texttt{uncertain} verdicts. Bold marks the errors that each ablation adds.}
\label{tab:ablation}
\small
\setlength{\tabcolsep}{4.5pt}
\begin{tabular}{llrrrrrr}
\toprule
Set & Configuration & TP & FP & TN & FN & U & Accuracy \\
\midrule
D2A-libtiff & Complete & 4 & 4 & 38 & 1 & 1 & 87.50\% \\
 & Without pre-slicing & 4 & \textbf{16} & 25 & 0 & 3 & 60.42\% \\
 & Without admission & 5 & \textbf{14} & 29 & 0 & 0 & 70.83\% \\
\midrule
SATE-V & Complete & 21 & 0 & 4 & 1 & 1 & 92.59\% \\
 & Without pre-slicing & 17 & 0 & 4 & 1 & \textbf{5} & 77.78\% \\
 & Without admission & 20 & 0 & 4 & \textbf{3} & 0 & 88.89\% \\
\bottomrule
\end{tabular}
\end{table}

Without pre-slicing, all-report accuracy falls from 87.50\% to 60.42\% on D2A-libtiff and from 92.59\% to 77.78\% on SATE-V. Lacking the initial context, the model promotes 16 D2A-libtiff false alarms, compared with four, and leaves five SATE-V reports \texttt{uncertain}, compared with one. Because the configuration without admission keeps the structured fact output, its errors isolate the effect of checking cited premises. Unsupported TP facts then promote 14 D2A-libtiff false alarms, and unsupported FP facts reject three SATE-V real defects, compared with four and one under admission. Accuracy falls to 70.83\% and 88.89\%. Admission thus blocks unsupported facts in both directions.

\takeaway{2}{Pre-slicing and admission both reduce errors, by supplying the context that the model lacks and by keeping unsupported facts from deciding in either direction.}

\subsection{RQ3: Reproducible Verdicts}
\label{sec:eval-llm4sa}

We compare \textsc{VulnSAGE} with LLM4SA, which inspects static warnings by asking an LLM questions enriched with domain knowledge and representative cases~\cite{wen2024llm4sa}. Both systems adjudicate each report of the four sets. The positive verdicts of each system, correct or not, enter a common PoC construction and execution procedure with the context that the system produced. Because targeted retrieval is part of \textsc{VulnSAGE}, it serves the PoC requests of \textsc{VulnSAGE}. Juliet enters the false-alarm comparison, and its synthetic cases need no PoC execution.

\begin{table}[t]
\centering
\caption{Comparison with LLM4SA. False alarms are shown as rejected over reference false alarms. Reproduced positives are positive verdicts that the common PoC procedure reproduces at runtime.}
\label{tab:llm4sa-comparison}
\small
\setlength{\tabcolsep}{5pt}
\begin{tabular}{lcccc}
\toprule
& \multicolumn{2}{c}{False alarms rejected} & \multicolumn{2}{c}{Reproduced positives} \\
\cmidrule(lr){2-3}\cmidrule(l){4-5}
Dataset & \textsc{VulnSAGE} & LLM4SA & \textsc{VulnSAGE} & LLM4SA \\
\midrule
Juliet & 47/50 & 44/50 & n/a & n/a \\
D2A-libtiff & 38/43 & 39/43 & 3 & 0 \\
SATE-V & 4/4 & 3/4 & 16 & 3 \\
ARVO Gold & n/a & n/a & 6 & 0 \\
\midrule
Total & 89/97 & 86/97 & 25 & 3 \\
\bottomrule
\end{tabular}
\end{table}

Both systems reject most false alarms, 89 of 97 for \textsc{VulnSAGE} and 86 for LLM4SA. On the three real-world sets, they reach 77 and 75 correct decisions out of 87. Their positive verdicts differ in what they lead to. Under the common PoC procedure, 25 positive verdicts of \textsc{VulnSAGE} reproduce at runtime, compared with three of LLM4SA. Admitted \textsc{VulnSAGE} facts record the carrier and the conditions that reach the reported operation, and targeted retrieval supplies the input format and invocation context that PoC construction requests. LLM4SA's context supports a verdict but seldom leads to an input that reaches the reported operation.

\takeaway{3}{\textsc{VulnSAGE}'s positive verdicts come with the program context that runtime reproduction needs.}

\subsubsection{Case Study: ARVO Finding 46653}
\label{sec:case-study-arvo-46653}

The ARVO finding numbered 46653 illustrates this context before PoC
construction.  The report points to an out-of-bounds access in OpenSC, but the
path value indexed at the access originates in a different function.  Over
PSEG, \textsc{VulnSAGE} follows the data flow of this path across function
boundaries.  It connects \texttt{get\_object\_path\_from\_object}, which obtains
the path from \texttt{obj->data}, to \texttt{sc\_pkcs15init\_delete\_object},
which passes the path through the file-selection check
\texttt{sc\_select\_file} and into \texttt{sc\_pkcs15init\_delete\_by\_path}.
The latter performs further file selection, access-control lookup and
authentication before reaching the reported operation, the
indexed access \texttt{file\_path->value[file\_path->len - 2]}.  The two
contexts contain the following call chains:

{\small\thickmuskip=4mu
\[
\begin{array}{@{}l@{\enspace}l@{}}
\textrm{\textsc{VulnSAGE}:} &
\texttt{get\_object\_path\_from\_object}
\rightarrow \texttt{sc\_pkcs15init\_delete\_object}
\rightarrow \texttt{sc\_select\_file} \\
&\rightarrow \texttt{sc\_pkcs15init\_delete\_by\_path}
\rightarrow \{\texttt{sc\_select\_file},\ \texttt{sc\_file\_get\_acl\_entry}, \\
&\phantom{\rightarrow{}} \texttt{sc\_pkcs15init\_authenticate}\}
\rightarrow \textit{reported operation} \\
\textrm{LLM4SA:} &
\texttt{sc\_pkcs15init\_delete\_object}
\rightarrow \texttt{sc\_pkcs15init\_delete\_by\_path}
\rightarrow \textit{reported operation}
\end{array}
\]
}

The braces group the file-selection, access-control and authentication
checks that the \textsc{VulnSAGE} context retains, shown at function
granularity.  They do not directly carry \texttt{file\_path->len}, but they
determine whether the object is selected as an elementary file and whether
execution reaches the reported operation.  The context omits the intermediate
logging call \texttt{sc\_print\_path}, which formats a message without carrying
the path to the reported operation.  The admitted source-to-sink fact cites relations from the object
path through the call argument and the receiving parameter to the indexed
access, which is the target of the fact.  When \texttt{file\_path->len} is one,
the index evaluates to $-1$, exposing the out-of-bounds read.

In the context that LLM4SA constructs for the same report,
\texttt{get\_object\_path\_from\_object}, \texttt{sc\_select\_file},
\texttt{sc\_file\_get\_acl\_entry} and \texttt{sc\_pkcs15init\_authenticate}
are absent.  This context therefore contains the deletion call chain without
the origin of the path or the file-selection and authentication conditions
under which execution reaches the reported operation.  This
case shows why dependency-guided context is more useful for PoC construction:
it preserves both the carrier and the conditions needed to explain and reach
the reported operation.  For this finding, \textsc{VulnSAGE} used that context
to construct an input with a one-element path, satisfy the file-selection and
authentication conditions, and trigger the out-of-bounds read.  The LLM4SA
context lacks the path-producing helper and these checks, and its PoC
attempt did not trigger the out-of-bounds read.  This finding is one of the
six ARVO reproductions of \textsc{VulnSAGE} in
Table~\ref{tab:llm4sa-comparison}, in which LLM4SA has no reproduced
positive on ARVO Gold.

\subsection{RQ4: Open-Source Validation}
\label{sec:eval-open-source}

We apply \textsc{VulnSAGE} to SAST scans of the 200 most-starred C/C++ repositories on GitHub, each at a fixed revision. Five SAST tools produce the alerts, namely Infer, Cppcheck, the Clang Static Analyzer (CSA)~\cite{clangsa}, CodeQL~\cite{codeql} and Semgrep~\cite{semgrep}. We keep the 45,440 alerts in the five supported families (CWE-121/122, CWE-401, CWE-416, CWE-457 and CWE-476). Each family includes closely related CWEs, such as out-of-bounds reads and writes for buffer overflow, double frees for invalid lifetime and uninitialized memory for uninitialized state. \textsc{VulnSAGE} adjudicates each alert and seeks a runtime reproduction for each \texttt{TP-S}.

\subsubsection{Scan Results}

Table~\ref{tab:rq4-scan} follows the alerts of each SAST tool through \textsc{VulnSAGE}. Static adjudication assigns \texttt{TP-S} to 1,040 of the 45,440 alerts and filters out the other 97.71\%. The filtering rate stays between 96.71\% and 99.09\% across the five tools, although their alert volumes differ by a factor of seven. Merging \texttt{TP-S} alerts that share a root cause yields 486 candidate defects, which reduces the review workload from 45,440 alerts to 486 defects.

PoC generation advances 104 of the 154 candidates whose targets run in our container to \texttt{TP-D}, a rate of 67.53\%. The other 332 candidates target execution environments beyond this container and keep their \texttt{TP-S} verdicts. All five SAST tools contribute reproduced defects, including CSA, CodeQL and Semgrep, which the labeled sets do not use. The reproduced defects span all five families, with 55 null dereferences, 31 buffer overflows, nine uses of uninitialized state, six lifetime errors and three memory leaks.

\begin{table}[t]
\centering
\caption{Open-source alerts in the five supported families and their outcomes. \texttt{TP-S} counts alerts. \texttt{TP-D} counts reproduced defects after merging alerts that share a root cause.}
\label{tab:rq4-scan}
\small
\setlength{\tabcolsep}{4pt}
\begin{tabular}{lrrrr}
\toprule
SAST tool & Alerts & \texttt{TP-S} & Filtered & \texttt{TP-D} \\
\midrule
Infer & 15,935 & 451 & 97.17\% & 47 \\
Cppcheck & 17,149 & 321 & 98.13\% & 14 \\
CSA & 6,135 & 121 & 98.03\% & 21 \\
CodeQL & 3,802 & 125 & 96.71\% & 18 \\
Semgrep & 2,419 & 22 & 99.09\% & 4 \\
\midrule
Total & 45,440 & 1,040 & 97.71\% & 104 \\
\bottomrule
\end{tabular}
\end{table}

\subsubsection{Upstream Confirmation}

Maintainers have so far responded to 26 reproduced defects, confirming 24 (92.31\%) and declining two as intended behavior. Table~\ref{tab:rq4-new} lists the ten confirmed defects that were previously unknown. These ten defects come from all five SAST tools. Maintainers fixed nine of them and accepted our patch for the tenth. They also handle three of them, in libtiff, radare2 and Deskflow, as security vulnerabilities. Because these ten defects had no public report or fix before ours, the training data of the model could not contain their labels or fixes. Fourteen further confirmed defects match issues that upstream already tracks in 11 projects, including Git, FlatBuffers, LZ4, Apache Arrow, and VLC.

\begin{table}[t]
\centering
\caption{Previously unknown defects that maintainers confirmed, grouped by SAST tool.}
\label{tab:rq4-new}
\small
\setlength{\tabcolsep}{4pt}
\begin{tabular}{llll}
\toprule
Project & Family & SAST tool & Upstream outcome \\
\midrule
libtiff & Buffer overflow & Infer & Fixed \\
radare2 & Null dereference & Infer & Fixed \\
Dragonfly & Null dereference & Cppcheck & Fixed \\
lwIP & Buffer overflow & Cppcheck & Patch accepted \\
Git & Null dereference & CSA & Fixed \\
Universal Ctags & Uninitialized state & CSA & Fixed \\
Binutils & Invalid lifetime & CSA & Fixed \\
7-Zip-zstd & Invalid lifetime & CSA & Fixed \\
LibRaw & Buffer overflow & CodeQL & Fixed \\
Deskflow & Buffer overflow & Semgrep & Fixed, high-severity security advisory \\
\bottomrule
\end{tabular}
\end{table}

\takeaway{4}{On real projects, \textsc{VulnSAGE} turns SAST alerts into reproduced defects that maintainers confirm, including previously unknown ones.}

We next trace two of these defects, in libtiff and 7-Zip-zstd, through static adjudication and runtime reproduction.

\subsubsection{Case Study: libtiff 4.5.0}
\label{sec:case-study-libtiff54}

This case study traces the report from static evidence construction to runtime reproduction and upstream confirmation. The SAST report identifies a buffer-overrun candidate in \texttt{tiff2pdf}. Starting from the report location, the report-centered slice recovers the numeric-conversion chain shown in Figure~\ref{fig:libtiff54-vulnerable-code}, whose metadata read lies several statements before the reported operation. In this chain, \texttt{StripByteCounts} is read through a \texttt{uint64\_t} carrier, a value is assigned to the narrower \texttt{uint32\_t} \texttt{max\_striplength}, that value sizes \texttt{stripbuffer}, and a later raw-strip read writes into the buffer. CLAIM exposes the corresponding carrier, allocation, and copy relations for structural admission, and the admitted fact supports a \texttt{TP-S} verdict.

\begin{figure}[t]
\begin{lrbox}{\fpmpanelboxA}\begin{minipage}[t]{\fpmhalfinner}
\begin{lstlisting}[style=fpmpanel,emph={sbc,max_striplength,stripbuffer}]
uint64_t *sbc;
uint32_t max_striplength = 0;
...
TIFFGetField(input,
    TIFFTAG_STRIPBYTECOUNTS, &sbc);
for (i = 0; i < stripcount; i++) {
    if (sbc[i] > max_striplength)
        // narrowing
        max_striplength = sbc[i];
}
\end{lstlisting}
\end{minipage}\end{lrbox}%
\begin{lrbox}{\fpmpanelboxB}\begin{minipage}[t]{\fpmhalfinner}
\begin{lstlisting}[style=fpmpanel,emph={sbc,max_striplength,stripbuffer}]
tsize_t striplength = 0;
...
stripbuffer = (unsigned char *)_TIFFmalloc(
    max_striplength);
...
striplength = TIFFReadRawStrip(
    input, i, (tdata_t)stripbuffer, -1);
\end{lstlisting}
\end{minipage}\end{lrbox}%
\fpmpanelpair{(a) Metadata read and narrowing}{(b) Allocation and raw-strip copy}
\caption{Relevant libtiff 4.5.0 source pattern. The report-level question is whether a large strip count can be narrowed before allocation while the later raw-strip read retains the larger effective length. Highlighted identifiers carry the strip length.}
\label{fig:libtiff54-vulnerable-code}
\Description{A 64-bit StripByteCounts value is assigned to a 32-bit maximum strip length used to allocate a buffer. A subsequent TIFFReadRawStrip call writes the strip into this allocation, exposing the narrowing and later-write chain.}
\end{figure}

PoC generation carries this static hypothesis into an AddressSanitizer-instrumented~\cite{serebryany2012asan} build of libtiff 4.5.0. The generated BigTIFF input sets \texttt{StripByteCounts} to \texttt{0x100000005}. On the vulnerable path, this value narrows to \texttt{5} when assigned to \texttt{max\_striplength}, so the program allocates a five-byte strip buffer. The later \texttt{TIFFReadRawStrip} call copies the raw strip into that buffer and AddressSanitizer reports a heap-buffer-overflow write of 4,294,967,301 bytes. The triggering command uses the \texttt{tiff2pdf} memory-limit option required to reach the vulnerable path; with that execution setting, the input produces a target-correlated sanitizer observation and advances the report from \texttt{TP-S} to \texttt{TP-D}.

The case study demonstrates the central design choices of \textsc{VulnSAGE}. The decisive evidence spans metadata retrieval, numeric conversion, allocation, and a later copy rather than a single reported line. The admitted static fact supplies a report-level TP basis, while PoC generation carries its trigger conditions into a concrete input and invocation. The correlated sanitizer hit provides an executable witness that maintainers can reproduce under the specified configuration, and the upstream developers subsequently confirmed and fixed the vulnerability.

\subsubsection{Case Study: 7-Zip-zstd}
\label{sec:case-study-7zip}

The second case covers a different analyzer and vulnerability family. CSA reports a use-after-free candidate in the extraction routine of a single-stream decoder, shown in Figure~\ref{fig:7zip-vulnerable-code}. The routine creates an output-stream object and keeps two pointers to it, a raw pointer and a reference-counted pointer that holds the sole counted reference. After decoding, the routine releases the counted reference. On the success branch, it then reads the output size through the raw pointer.

\begin{figure}[t]
\begin{lrbox}{\fpmpanelboxA}\begin{minipage}[t]{\fpmhalfinner}
\begin{lstlisting}[style=fpmpanel,language=C++,emph={outStreamSpec,outStream}]
CDummyOutStream *outStreamSpec =
    new CDummyOutStream;
CMyComPtr<ISequentialOutStream>
    outStream(outStreamSpec);
...
for (;;) {
    result = decoderSpec->CodeResume(
        outStream, &unpackedSize, progress);
    ...
}
\end{lstlisting}
\end{minipage}\end{lrbox}%
\begin{lrbox}{\fpmpanelboxB}\begin{minipage}[t]{\fpmhalfinner}
\begin{lstlisting}[style=fpmpanel,language=C++,emph={outStreamSpec,outStream}]
decoderSpec->ReleaseInStream();
// drops the last counted reference
outStream.Release();
...
else if (result == S_OK) {
    // read after free
    _unpackSize = outStreamSpec->GetSize();
    ...
}
\end{lstlisting}
\end{minipage}\end{lrbox}%
\fpmpanelpair{(a) Object creation and decoding}{(b) Release and size read}
\caption{Relevant 7-Zip-zstd Zstandard-handler code. The report-level question is whether the release frees the object before the success branch reads its size through the raw pointer. Highlighted identifiers refer to the output-stream object.}
\label{fig:7zip-vulnerable-code}
\Description{A raw pointer and a reference-counted pointer refer to an output-stream object. The counted reference is released after decoding, and the success branch later reads the object's size through the raw pointer.}
\end{figure}

The report-centered slice from the reported read follows the raw pointer back to its allocation and to the construction of the counted pointer, and its control-flow relations place the release call between them. Freeing the object on release follows from a reference-counting convention that program relations do not express. CLAIM assigns this interpretation to the model, which states as a fact that releasing the last counted reference deletes the object. The fact cites three relations that program analysis checks: the binding of both pointers to the object allocated in the routine, the release call, and the later member read through the raw pointer on the success branch. With these premises admitted, the fact supports a \texttt{TP-S} verdict for the invalid-lifetime family.

PoC generation needs no malformed input for this report, because valid Zstandard files from the standard compressor reach the success branch. Under AddressSanitizer, testing such a file with the archive tool reports an eight-byte heap-use-after-free read at the size query in each run, which advances the report to \texttt{TP-D}. The uninstrumented build reports success on this input because the freed block still holds the old size. Ordinary tests therefore miss the defect. The maintainers fixed it the day after our report by moving the release below the size query in all five single-stream handlers that share this code.

\section{Related Work}
\label{sec:related}

\paragraph{LLM-assisted report adjudication.}
LLift investigates use-before-initialization warnings with post-constraint-guided reasoning and function summaries, then asks the model to reassess its conclusions~\cite{llift2023}. LLM4SA asks LLMs questions enriched with domain knowledge and representative cases~\cite{wen2024llm4sa}. ZeroFalse enriches alerts with flow traces and CWE-specific knowledge~\cite{zerofalse2025}, and an industrial study compares such techniques on enterprise alarms~\cite{du2026empirical}. These works aim to classify alerts accurately. \textsc{VulnSAGE} targets the basis of each verdict. It decides reports from facts whose cited program relations hold, and supported reports proceed to runtime reproduction.

\paragraph{Integration of LLMs and program analysis.}
IRIS uses LLMs to label APIs as taint sources and sinks for CodeQL taint queries, then asks the LLM to classify the resulting alerts in their code context~\cite{iris2024}. GPTScan asks GPT to identify key variables and statements for each smart-contract vulnerability type and confirms them with a static check that the type prescribes, such as data-flow tracing or order checking~\cite{gptscan2024}. InferROI applies resource-oriented intention inference to static resource-leak detection~\cite{inferroi2025}. In CLAIM, the model states a TP or FP interpretation of an existing report and selects the relations it assumes from a shared vocabulary. Admission checks these relations per fact, and the verdict follows from admitted facts.

\paragraph{Program slicing and representations.}
Slicing recovers statements relevant to a value at a program point~\cite{weiser1984slicing}. Dependence graphs extend slicing across procedures~\cite{horwitz1988interprocedural}. CPGs combine syntax with control and data-flow representations~\cite{yamaguchi2014cpg}. LLMxCPG fine-tunes a model to write Joern queries that extract vulnerability-focused slices, which a second model classifies~\cite{llmxcpg2025}. Our slicer derives its criteria from the report and follows PSEG relations that explain dependencies. Its output retains source identity, which lets CLAIM facts cite the code that the model reads.

\paragraph{Runtime reproduction and PoC generation.}
NodeMedic-FINE uses dynamic taint information to guide exploit synthesis for Node.js vulnerabilities~\cite{cassel2025nodemedic}. ProtocolGuard combines LLM-guided static analysis with dynamic verification, generating assertions and PoC tests for protocol non-compliance~\cite{song2026protocolguard}. SyzBridge adapts kernel PoCs across environments~\cite{zou2024syzbridge}. \citet{agrawal2026noise} add LLM-based exploit generation to a hybrid SAST triage pipeline. \textsc{VulnSAGE} guides reproduction from an existing SAST report and its admitted static TP basis.

\section{Limitations}
\label{sec:discussion}

\paragraph{Semantic interpretation.}
CLAIM links semantic facts to program premises. Admission checks the cited relationships, while the model interprets properties such as capacity or ownership validity. That interpretation can be incorrect even when its structural premises are supported. Semantic models of project contracts could extend direct checking to these properties.

\paragraph{Report and structural coverage.}
Coverage depends on the upstream reports and the PSEG relations available through the PSEG service for their matching source revisions. Macro expansion or indirect calls can leave program relationships unresolved, as can external-library behavior. These gaps can prevent retrieval gates or admission conditions from being satisfied. Broader frontend and relation support would extend the reports for which a decision is possible.

\paragraph{Model variability.}
Models and prompting configurations can produce different context requests or interpretations. Program-analysis pipelines can also be nondeterministic~\cite{miao2025nondeterminism}. Our results describe the evaluated configuration. Public projects and benchmarks may include material encountered during model training~\cite{ding2025codelm,llm4vuln2024}. Structural admission checks cited relations independently of the model, while prior exposure may affect its semantic judgments.

\paragraph{Runtime reproduction.}
Reproduction requires a buildable revision and an input that reaches the reported operation under a suitable configuration. SyzBridge documents the need to adapt kernel PoCs across environments~\cite{zou2024syzbridge}. Broader input construction and environment support could reproduce more statically supported reports. Each reproduction concerns its tested input and configuration.

\section{Conclusion}
\label{sec:conclusion}

\textsc{VulnSAGE} connects LLM reasoning to program structure in two directions. Report-centered slicing over PSEG gives the model the dependencies behind a report together with the structure that explains them. CLAIM returns the model's interpretations to program structure as semantic facts with premises that program analysis checks, and the verdict follows from facts whose premises hold. The resulting judgments guide tests that maintainers can inspect and rerun. On open-source projects, these reproductions include ten previously unknown defects confirmed by upstream maintainers, nine of which are already fixed.

\bibliographystyle{ACM-Reference-Format}
\bibliography{references}

\end{document}